\documentclass[aps,nofootinbib,twocolumn,floatfix,showpacs,preprintnumbers]{revtex4}
\usepackage{epsfig,amssymb}

\newcommand{\be}{\begin{equation}}
\newcommand{\ee}{\end{equation}}
\newcommand{\bea}{\begin{eqnarray}}
\newcommand{\eea}{\end{eqnarray}}
\newcommand{\gag}{g_{a\gamma}}

\newcommand\I{{\rm i}}
\def\lsim{\mathrel{\raise.3ex\hbox{$<$\kern-.75em\lower1ex\hbox{$\sim$}}}}
\def\gsim{\mathrel{\raise.3ex\hbox{$>$\kern-.75em\lower1ex\hbox{$\sim$}}}}
\newcommand\E{{\rm e}}
\newcommand\hs{{\bf \hat{s}}}
\newcommand\ha{{\bf  \hat{a}}}
\newcommand\hb{{\bf  \hat{b}}}

\def\d{{\rm d}}

\def\alt{\raise0.3ex\hbox{$\;<$\kern-0.75em\raise-1.1ex\hbox{$\sim\;$}}}
\def\agt{\raise0.3ex\hbox{$\;>$\kern-0.75em\raise-1.1ex\hbox{$\sim\;$}}}
\begin{document}
 \title{The Milky Way as a Kiloparsec-Scale Axionscope}
\author{Melanie Simet}
\affiliation{Department of Astronomy \& Astrophysics, The University of Chicago, Chicago, IL~~60637-1433}
\author{Dan Hooper and Pasquale D. Serpico}
\affiliation{Center for Particle Astrophysics, Fermi National
Accelerator Laboratory, Batavia, IL~~60510-0500}

\date{\today}
\begin{abstract}

Very high energy gamma-rays are expected to be absorbed by the extragalactic background light over cosmological distances via the process of electron-positron pair production. Recent observations of cosmologically distant gamma-ray emitters by ground based gamma-ray telescopes have, however, revealed a surprising degree of transparency of the universe to very high energy photons. One possible mechanism to explain this observation is the oscillation between photons and axion-like-particles (ALPs). Here we explore this possibility further, focusing on photon-ALP conversion in the magnetic fields in and around gamma-ray sources and in the magnetic field of the Milky Way, where some fraction of the ALP flux is converted back into photons. We show that this mechanism can be efficient in allowed regions of the ALP parameter space, as well as in typical configurations of the Galactic Magnetic Field. As case examples, we consider the spectrum observed from two HESS sources: 1ES1101-232 at redshift $z=0.186$ and H 2356-309 at $z=0.165$. We also discuss features of this scenario which could be used to distinguish it from standard or other exotic models.
\end{abstract}
\preprint{FERMILAB-PUB-07-658-A}
\maketitle

\maketitle

\section{introduction}
Very high energy (VHE, $E_{\gamma} \gsim 0.1$ TeV) gamma-rays are
expected to be attenuated over cosmological distances by infrared,
optical and ultraviolet photons of the extragalactic background
light (EBL) via the process of pair production ($\gamma_{\rm VHE} +
\gamma_{\rm EBL} \rightarrow e^+ e^-$). This mechanism is expected to strongly suppress the VHE spectra from distant sources and
opens the possibility of measuring the spectrum and density of the
EBL via gamma-ray observations~\cite{Stecker:1992wi}. Recent
findings by imaging atmospheric Cherenkov telescopes, however,
indicate a relatively large degree of transparency of the universe
to gamma-rays~\cite{Aharonian:2005gh,Mazin:2007pn}. Taken at face
value, the data seem to require a lower density of the EBL than expected
and/or considerably harder injection spectra than initially
thought~\cite{Stecker:2007jq,Stecker:2007zj}. On the other hand, it is also
possible that some exotic mechanism is responsible for the observed
lack of attenuation.

In Ref.~\cite{De Angelis:2007dy}, it was argued that the
oscillations of photons into axion-like particles (ALPs) in
extragalactic magnetic fields could provide a way of avoiding the
exponential suppression of the gamma-ray spectrum resulting from
pair production. This mechanism is especially effective for
extremely distant sources and in the presence of relatively strong
(nanogauss scale) extragalactic magnetic fields. While consistent
with present bounds, there is no observational evidence for
such strong fields, which appear hard to generate
dynamically. For example, recent simulations based on
magneto-hydrodynamic amplification of seed magnetic fields driven
by structure formation predict fractions of extragalactic space
containing nanogauss or stronger fields ranging from less than
$0.1\%$~\cite{Dolag:2004kp} to $10\%$~\cite{Sigl:2004gi}.
Also, recent results from the Pierre Auger
Observatory appear to favor magnetic fields which are closer to the
weaker estimates, with upper limits for Mpc coherence lengths  in the $0.1\div 1\,$nG range
in most of the sky~\cite{Collaboration:2007si,DeAngelis:2007rw}. In this respect, although photon-ALP oscillations in extragalactic
magnetic fields may in principle be possible, it would require unsupported assumptions on the fields in order to act as an effective mechanism for avoiding the attenuation of VHE gamma-ray sources.

A somewhat different approach was taken in
Refs.~\cite{Hooper:2007bq} and~\cite{Hochmuth:2007hk} where it was
argued that the magnetic fields present in astrophysical
accelerators themselves would lead to significant photon-ALP
conversion over a large range of ALP parameter space,
including a small region of the very interesting parameter space in
which Peccei-Quinn axions provide a solution to the strong CP
problem.  This effect is expected to become significant at gamma-ray
energies, including the range being explored by the ground based
gamma-ray telescopes HESS~\cite{HESSurl}, MAGIC~\cite{MAGICurl},
VERITAS~\cite{VERITASurl} and CANGAROO-III~\cite{CANGAROOIIIurl}, as
well as of the forthcoming satellite mission, GLAST~\cite{GLASTurl}.
In this article, we explore the possibility of reducing the
cosmological attenuation of VHE gamma-rays through a photon-ALP
mixing mechanism, but invoking only known magnetic fields: those
needed to confine and accelerate cosmic rays at the gamma-ray
sources, where a significant ALP  flux is produced  via
oscillations. To reconvert a fraction of the ALP flux back into photons, well after
the primary photon flux has disappeared due to the absorption on the
EBL, we rely only on the magnetic field of the Milky Way. This mechanism can lead to the appearance of a substantial flux of
gamma-rays (up to $\sim$30\% of the low-energy extrapolation) at
energies where none is expected to be seen. It has also some other
very peculiar observational predictions, which we will also discuss.

The remainder of this paper is structured as follows. In
Sec.~\ref{formalism} we describe the mechanism proposed. In
particular, we devote Sec.~\ref{MWbackconversion} to the details of
the reconversion of ALPs in the Galactic Magnetic Field. In
Sec.~\ref{atten} we describe how we treat the absorption of photons
by the EBL during the propagation from their sources to the Milky
Way.  In Sec.~\ref{results}, in order to illustrate the impact of
such mechanism in a realistic scenario, we present the reconstructed
spectrum for two HESS sources: 1ES1101-232 at redshift $z=0.186$ and
H 2356-309 at $z=0.165$.  In Sec.~\ref{conclusions} we discuss and
summarize our results, focusing on the distinctive observational
features of the scenario outlined here.

\section{The mechanism}\label{formalism}
As discussed in Refs.~\cite{Hooper:2007bq,Hochmuth:2007hk}, given
the typical sizes and magnetic field strengths present in
astrophysical accelerators, significant photon to ALP conversion can
occur in or near the VHE gamma-ray sources over a large range of allowed ALP parameter
space. In the limit of complete ``depolarization'' of the photon-ALP
system, one expects $1/3$ of the original photons to be converted
into ALPs at the source above a critical energy ${\cal E}\equiv m^2_{a}/(2\,g_{a\gamma}B)$,
where $m_a$ is the ALP mass, $g_{a\gamma}$ is the ALP-photon coupling and $B$ the
field strength.  That
is, for $E\gg{\cal E}$ there could potentially be an ALP flux from VHE
gamma-ray sources as large as $\sim$50\% of the residual gamma-ray flux
exiting the source.  The energy ${\cal E}$ naturally falls in the
gamma-ray band. Although it is not crucial for
the viability of the mechanism being discussed, we shall assume that 
${\cal E}\ll 0.1\,$TeV, thus insuring that the spectrum observed by ground based gamma-ray telescopes will not demonstrate peculiar spectral 
features resulting from the onset of photon-ALP oscillations. As we shall see
below, this condition is naturally fulfilled over the most viable range of parameters for which the mechanism is efficient (see~\cite{Hooper:2007bq} for details). Thus, over the whole range
of energy observed by atmospheric Cherenkov telescopes, the ALP
flux generated at the source via oscillation follows exactly the
same spectrum of the photons at the source. 

This photon-ALP mix then
propagates towards the Milky Way without further oscillations, since
in absence of an external field the ALPs and the photons effectively
decouple. During propagation over cosmological distances, the spectrum of
photons is depleted via pair production.  As a result, upon reaching the Milky Way, the flux will be dominated by ALPs, some of which can be converted back
into photons by the Galactic Magnetic Field. Since this step is
crucial and it constitutes the more original part of the mechanism
we propose, in Sec.~\ref{MWbackconversion} we describe it in detail.

\subsection{Reconversion in the Milky Way}\label{MWbackconversion}
In a given direction in Galactic coordinates, $(b,l)$, the unit vector, $\hs$, with components $\{\cos b\cos l,- \cos b\sin l,\sin b\}$ can be constructed. Through the usual orthogonalization procedure of Gram-Schmidt, $\hs$ can be completed into a $\mathbb{R}^3$ basis (along with, say, $\ha$ and $\hb$, two vectors orthogonal to $\hs$). One can thus decompose the magnetic field, ${\bf B}({\bf x})$, into components $\{B_a,B_b,B_s\}$ in this basis and show that the probability of an ALP converting into a photon (of any polarization) while traveling a distance $s$ along $\hs$ is given by~\cite{Mirizzi:2007hr}:
\begin{eqnarray}\label{BxBy}
 P_{a\to \gamma}(s)&=&\frac{\gag^2}{4}
 \bigg(\left|\int_0^s{\rm d}s^{\prime}
 \E^{{\I}(\Delta_a-\Delta_{\rm pl})\,s^{\prime}}B_{a}(s^{\prime})
 \right|^2+\nonumber\\
 &&\left |\int_0^s{\rm d}s^{\prime}
 \E^{{\I}(\Delta_a-\Delta_{\rm pl})\,
 s^{\prime}}B_{b}(s^{\prime})\right|^2\bigg)
\end{eqnarray}
This probability is a factor of 2 larger than $P_{\gamma\to a}(s)$ given in Ref.~\cite{Mirizzi:2007hr} due to the two photon states available for the ALP to convert into. In this expression, the quantities entering the phase factor are $\Delta_a = -m^2_{a}/2E$ and $\Delta_{\rm pl}= \omega_{\rm pl}^2/2\,E$, where $\omega_{\rm pl} =\sqrt{ 4\pi\alpha\,n_e/m_e}$ is the plasma frequency, $E$ the photon energy,  $m_e$ the electron mass,  $n_e$ the electron density, and $\alpha$ the fine structure constant.  For the following considerations, it is useful to introduce the following dimensionless quantities: $g_{11}=\gag/10^{-11}~{\rm GeV}^{-1}$,  $s_{\rm kpc}\equiv s/{\rm kpc}$,  $m_{\rm neV}\equiv m$/neV and $E_{\rm TeV}\equiv E/$TeV. Results from the CAST experiment~\cite{Andriamonje:2007ew} provide a direct bound on the ALP-photon coupling of $g_{11} \lesssim 8.8$ for $m_a\alt 0.02\,$eV, nominally below the long-standing globular cluster limit~\cite{acconseq}. For ultra-light ALPs ($m_a\alt 10^{-11}\,$eV), the absence of gamma-rays from SN~1987A yields a stringent limit of $g_{11}\lesssim 1$~\cite{Brockway:1996yr} or even $g_{11} \lesssim 0.3$~\cite{Grifols:1996id}. In the range $10^{-11}\,{\rm eV}\ll m_a\ll 10^{-2}\,$eV, the CAST bound is the most general and stringent. For a range of $\mu$eV-scale masses, bounds from ADMX~\cite{Bradley:2003kg} are stronger, but rely on the assumption that axions constitute the Galactic dark matter. In these illustrative units, the phases can be rewritten as:
\begin{equation}\label{eq12}
\begin{array}{ccc}
\Delta_a\, s_{\rm kpc}&=& -0.77\times 10^{-4}m_{\rm neV}^2\,s_{\rm kpc}\,E_{\rm TeV}^{-1}\,,\\
&&\\
\Delta_{\rm pl}\, s_{\rm kpc}&=& -1.11\times
10^{-7}\,\left(\frac{n_e}{{\rm cm}^{-3}}\right) s_{\rm kpc}\,E_{\rm TeV}^{-1}\,.
\end{array}
\end{equation}

The integrals in Eq.~(\ref{BxBy}) are very small unless the phases of the integrands vanish over typical Galactic sizes of $\sim 10$ kpc. Barring fine-tuning, we must require $|\Delta_i \times 10\,{\rm kpc}|\ll1$, leading to
\begin{equation}\label{eq12}
\begin{array}{ccc}
m_{\rm neV}&\ll& \sqrt{1.3\times 10^3\, E_{\rm TeV}}\,,\\
&&\\
n_e&\ll& 0.90\times 10^6\,{\rm cm^{-3}}
\,E_{\rm TeV}\,.
\end{array}
\end{equation}

For the energies of interests for atmospheric Cherenkov telescopes,
the second condition is always satisfied in the Milky Way. The first
condition only holds if $m_a\ll10^{-7}\,$eV.  The optimal range of
parameters for this mechanism is thus $10^{-10}\,{\rm eV}\alt m_a\alt
10^{-8}\,{\rm eV}$, over which couplings as large as $g_{11}\approx 8$ are
consistent with present bounds.

\begin{figure}[!tb]
\begin{center}
\begin{tabular}{c}
\epsfig{file=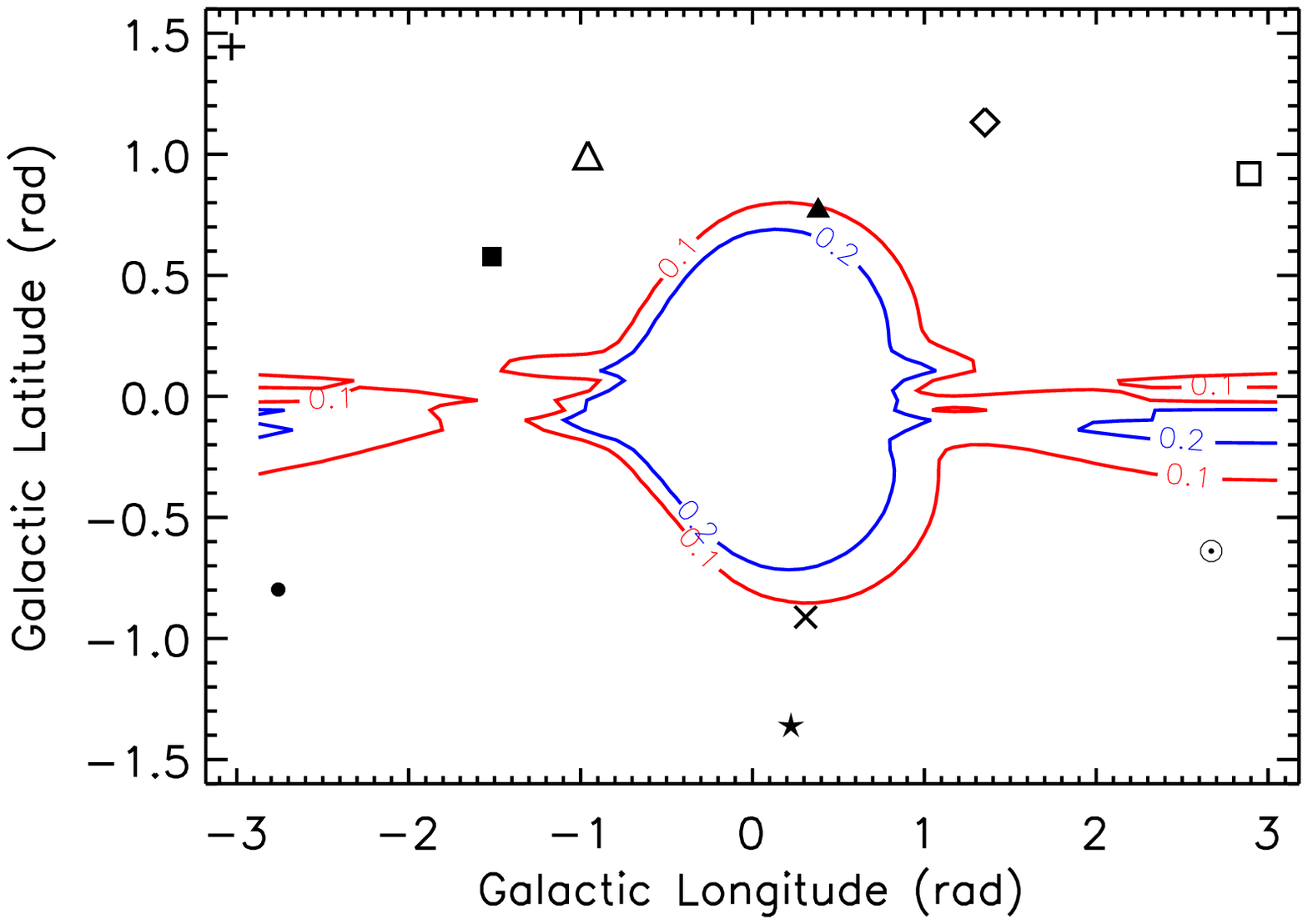,width=0.98\columnwidth}\\
\epsfig{file=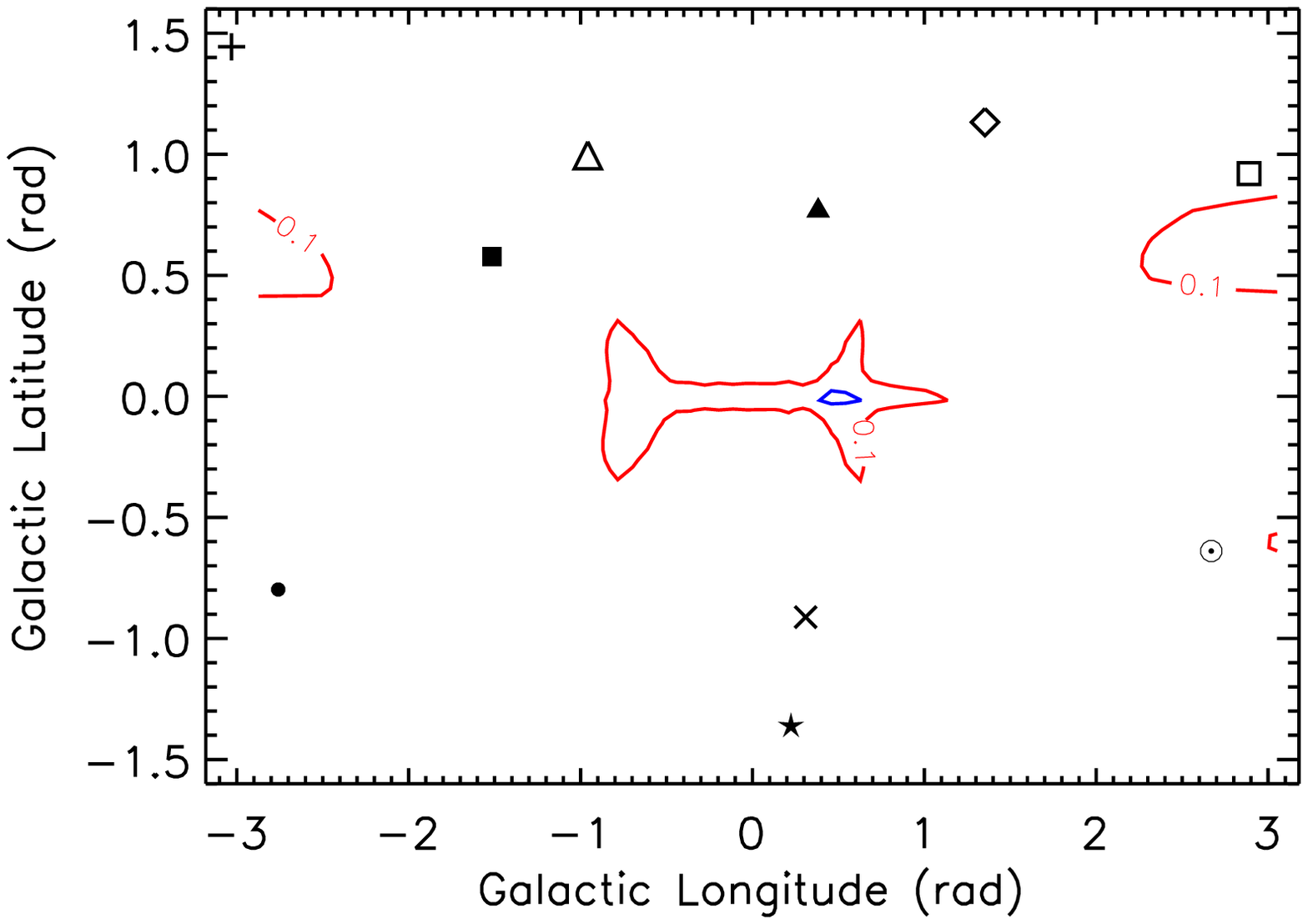,width=0.98\columnwidth}\\
\epsfig{file=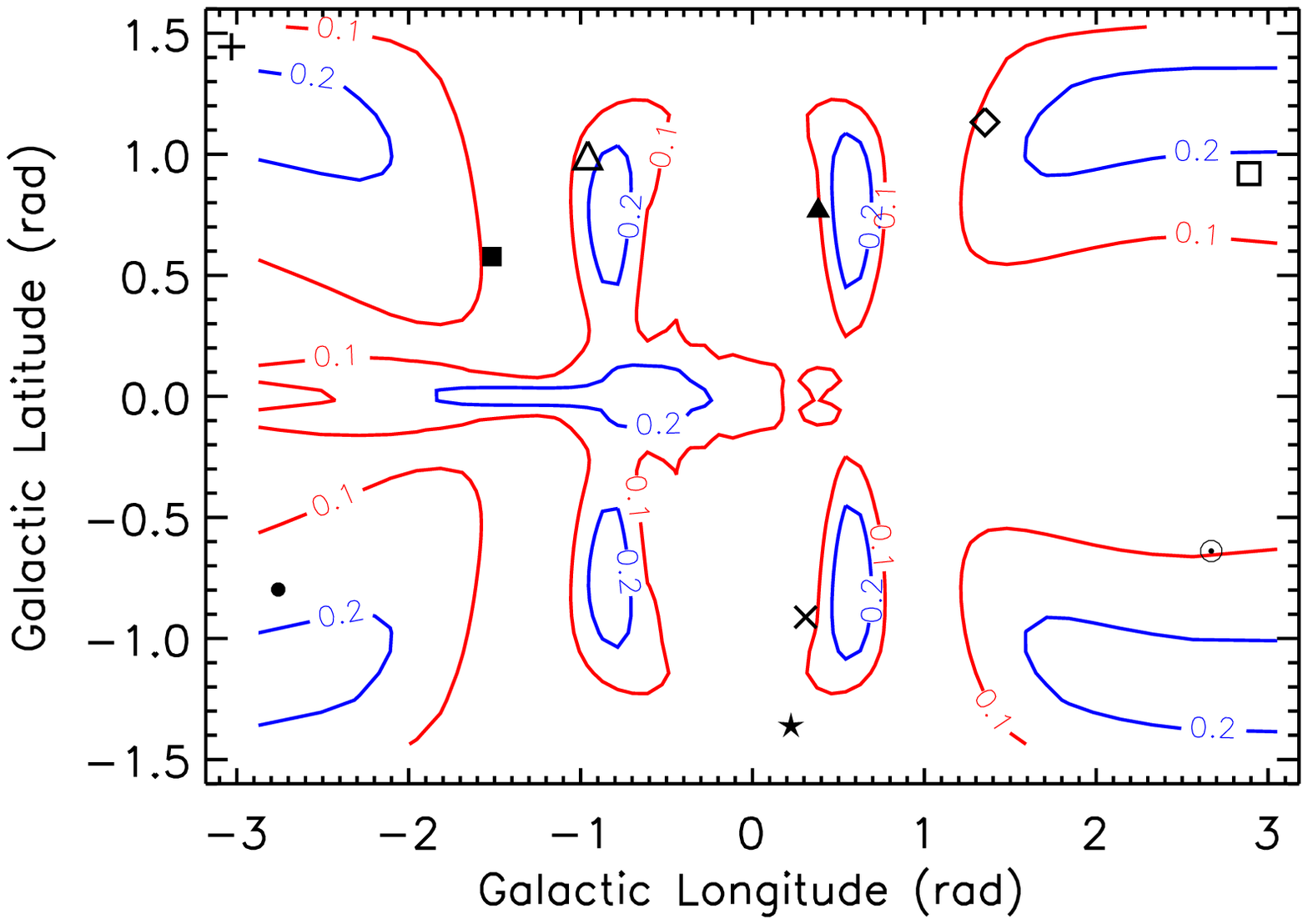,width=0.98\columnwidth}
\end{tabular}
\end{center}
\caption{The probability of ALP-photon conversion in three different models of the Galactic Magnetic Field (PS, TT, HMR from top to bottom), for $g_{11}=5$ and assuming that the conditions of Eq.~(\ref{eq12}) are satisfied.
  Also shown are the locations of the known very high energy gamma-ray sources at  redshift greater than 0.1, labeled according to the symbols found in table I.\label{GMFmaps}}
\end{figure}

The conversion probability also depends on the geometry of the
Galactic Magnetic Field, which is not very well known, especially in
the directions toward the Galactic Center and away from the Galactic
Plane. In Fig.~\ref{GMFmaps}, we show isocontours in ALP-photon
conversion probability for three different models of the Galactic
Magnetic Field (see Ref.~\cite{Kachelriess:2005qm}). Following the
authors' initials, we label the models as PS~\cite{Prouza:2003yf},
TT~\cite{Tinyakov:2001ir}, and HMR~\cite{Harari:1999it}. Physically,
the PS model includes a dipolar field which produces a small
component perpendicular to the Galactic Plane in the Solar
neighborhood but is responsible for a strong field in the Galactic
Bulge. The PS model also includes a toroidal halo field at kpc
distances above and below the Galactic Plane. The HMR model does not
include a dipolar field, but has a more prominent field in the halo
characterized by the same symmetry pattern of the field in the disk.
Finally, the TT model is the most conservative of the three,
with a weaker intensity for the local coherent field and no
prominent regular field in the Galactic Center or in the halo. Each
of the models implement a thin disk field following the spiral
pattern of the Galaxy, but the symmetry with respect to the Galactic
Plane in the TT model is opposite to the one assumed in HMR and PS.
Further details can be found in Ref.~\cite{Kachelriess:2005qm}.

The isocontours of ALP-photon conversion probability shown in
Fig.~\ref{GMFmaps} were calculated using Eq.~(\ref{BxBy}), for
$g_{11}=5$ and assuming that the conditions of Eq.~(\ref{eq12}) are
satisfied. This means that the probability can be rescaled according
to $g_{11}^2$, as long as $ P_{a\to \gamma}\ll 1$ (Eq.~(\ref{BxBy})
indeed only holds at leading order in perturbation theory).  In the
limit of full mixing in the Galaxy, one would obtain a
reconversion probability of 2/3.

For illustration, we also show in Fig.~\ref{GMFmaps} the positions
of all currently known VHE gamma-ray sources at a redshift of 0.1 or
greater, symbol-coded according to Table~I. It is clear that for
$g_{11}=5$ there are regions of the sky in which the reconversion
probability can be 20\% or even larger, and that a significant
fraction of the sky corresponds to a probability larger than 10\%.
It is intriguing to notice that, in the HMR model, many of the VHE
gamma-ray sources lie within or nearby these regions. We conclude
that appreciable reconversion probabilities are possible, although
difficult to predict given the scarcity of our knowledge regarding the
structure of the Galactic Magnetic Field. In the
remaining of this paper, we shall assume reconversion probabilities of
$\sim$0.1 and explore the resulting phenomenological consequences.
Note that for $g_{11}\sim 5$, $m_a\sim10^{-9}\,$eV and microgauss-scale
fields, the critical energy ${\cal E}$ falls in the sub-GeV range, consistently with
the assumption put forth in Sec. II.

\begin{table}[!htb]
\begin{tabular}{|c||c|c|c|c|c|c|}
\hline
Object       & $z$ & $l$ &  $b$ & Ref. & Symbol\\
 \hline
 \hline
3C279 & 0.536     & 305.10  &    57.06 & \cite{Teshima:2007zw} & $\triangle$\\
\hline
 PG 1553+113 & $>0.25 $     &  21.91     & 43.96 & \cite{PG1553} &$\blacktriangle$\\
\hline
 1ES1011+496 & $0.212$     &  165.53  &  52.71 & \cite{Albert:2007kw} & $\square$\\
\hline
1ES 0347-121 & 0.188 &   201.93 &  -45.71 & \cite{Aharonian:2007tc}  & $\bullet$\\
\hline
1ES1101-232 & 0.186 & 273.19   & 33.08 & \cite{1ES1101} & $\blacksquare$\\
\hline
1ES1218+304  & 0.182 &  186.36 &     82.73 & \cite{1ES1218} & $+$\\
\hline
H 2356-309 & 0.165 &  12.84   & -78.04  & \cite{H2356}& $\star$\\
\hline
1ES 0229+200 &   0.140 &  152.94 & -36.61 & \cite{Aharonian:2007wc} & $\odot$\\
\hline
 H 1426+428 & $0.129 $  &  77.49 & 64.90  & \cite{H1426}& $\diamond$\\
\hline
PKS 2155-304 & 0.117  &  17.73  & -52.25 & \cite{PKS2155}& $\times$\\
\hline
   \end{tabular}
\caption{The current catalog of known very high energy gamma-ray sources at redshift greater than 0.1. The locations of these objects are also shown in Fig.~\ref{GMFmaps}.\label{sources}}
\end{table}
%
\subsection{Photon absorption onto the EBL}\label{atten}
To study the quantitative predictions of our model and make comparisons with the available data, we must also account for the standard dimming of the VHE photons from cosmologically distant sources. The spectra of distant VHE gamma-ray sources will be attenuated through electron-positron pair production.  This attenuation takes the form $\exp(-\tau)$, where $\tau(E,z_0)$ is given by the equation:

\begin{equation}
\tau= \int_0^{z_0} \frac{\d z}{(1+z)H(z)}\int \d\omega \frac{\d n}{\d\omega}(\omega,z)
\,\bar\sigma(E,\omega,z)\,,
\end{equation}
where
\begin{equation}
\bar\sigma  (E,\omega,z)   \equiv \int_{-1}^{1-2/E\,\omega}\frac{1}{2} (1-\mu)\sigma_{\gamma\gamma}(E,\omega,\mu)\,\d\mu \,.
\end{equation}
Here, $z_0$ is the source redshift, $H(z)$ is the rate of Hubble
expansion, $E$ and $\omega$ are the (appropriately redshifted)
source and background photon energies, respectively, $\mu$ is the
cosine of the angle between the incoming and target photon, and
$\sigma_{\gamma\gamma}$ is the cross section for electron-positron
pair production~\cite{steckerbook}.

\begin{figure}
\begin{center}
\epsfig{file=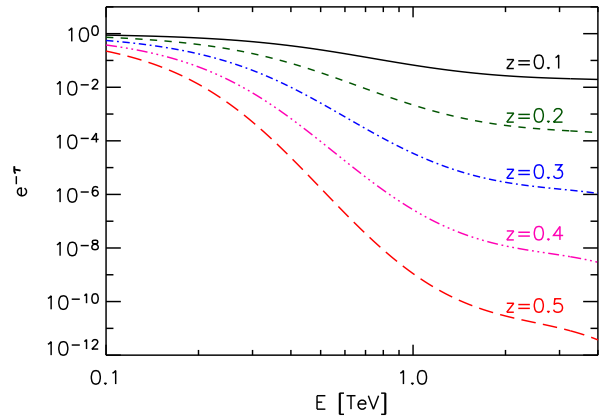,width=1.0\columnwidth, angle=0}
\end{center}
\caption{Suppression of the gamma-ray spectrum due to pair production for sources at redshifts of $z=0.1$, $0.2$, $0.3$, $0.4$, and $0.5$, from top to bottom.\label{suppression}}
\end{figure}

The distribution of EBL photons in energy and redshift which we use, $\d n/
\d\omega$, is based on the integrated galaxy light model of Ref.~\cite{primack} and the data of
Ref.~\cite{hauserdwek}.  We assume
that the shape and normalization of the background radiation does
not change except for redshifting between the source and the
observer, which as suggested by the models of Ref.~\cite{kneiske} is
a reasonable assumption for $z<0.4$. For the sources being
considered here (see table I), we expect this to be a reasonable assumption.
In Fig.~\ref{suppression}, we plot the attenuation as a function of
observed energy for sources at various redshifts.  Note
that it is not only the overall intensity of the source but the
shape of the spectrum that is affected by this suppression.

\subsection{Results}\label{results}

\begin{figure}[!htb]
\begin{tabular}{c}
\epsfig{file=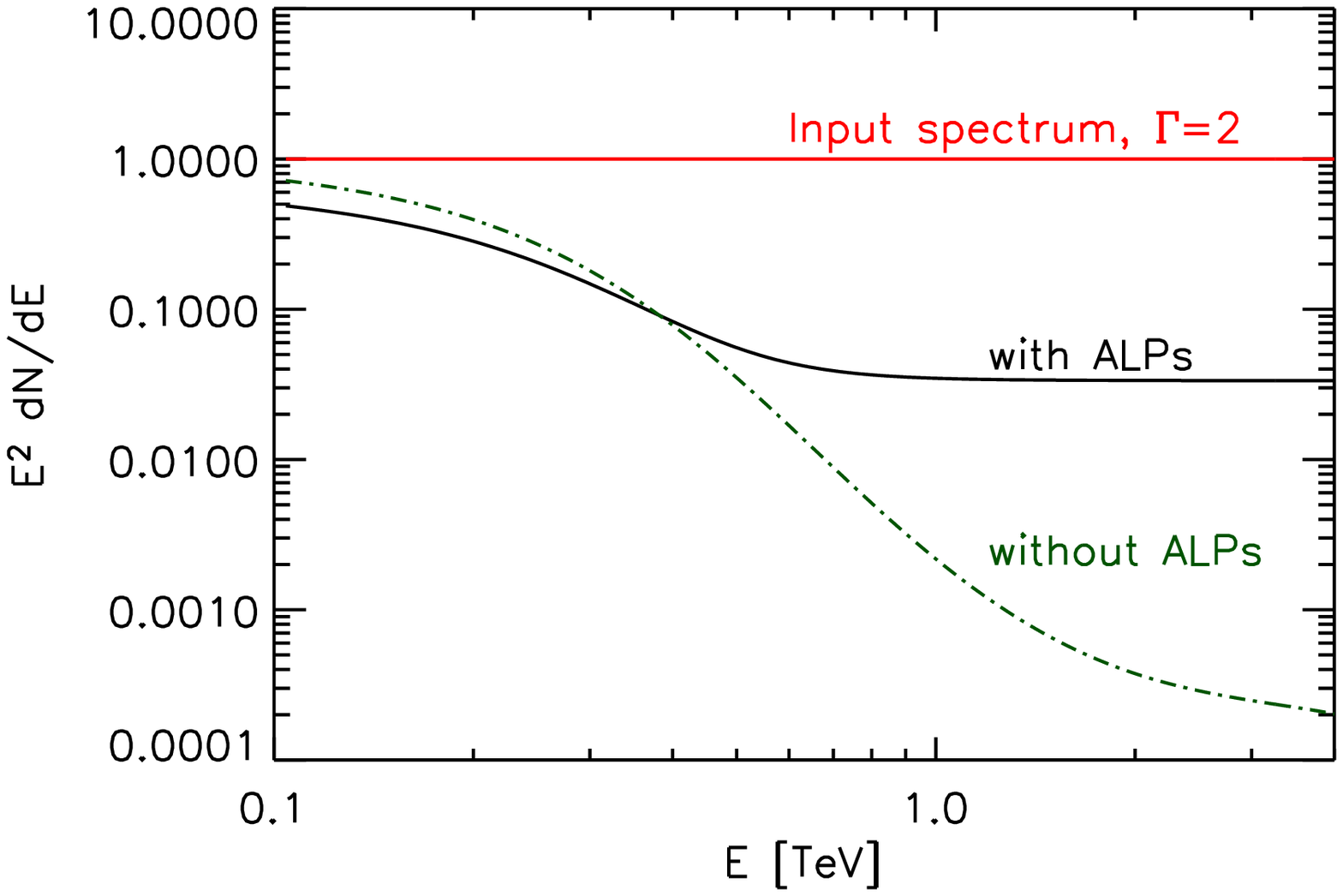,width=1.0\columnwidth, angle=0} \\
 \epsfig{file=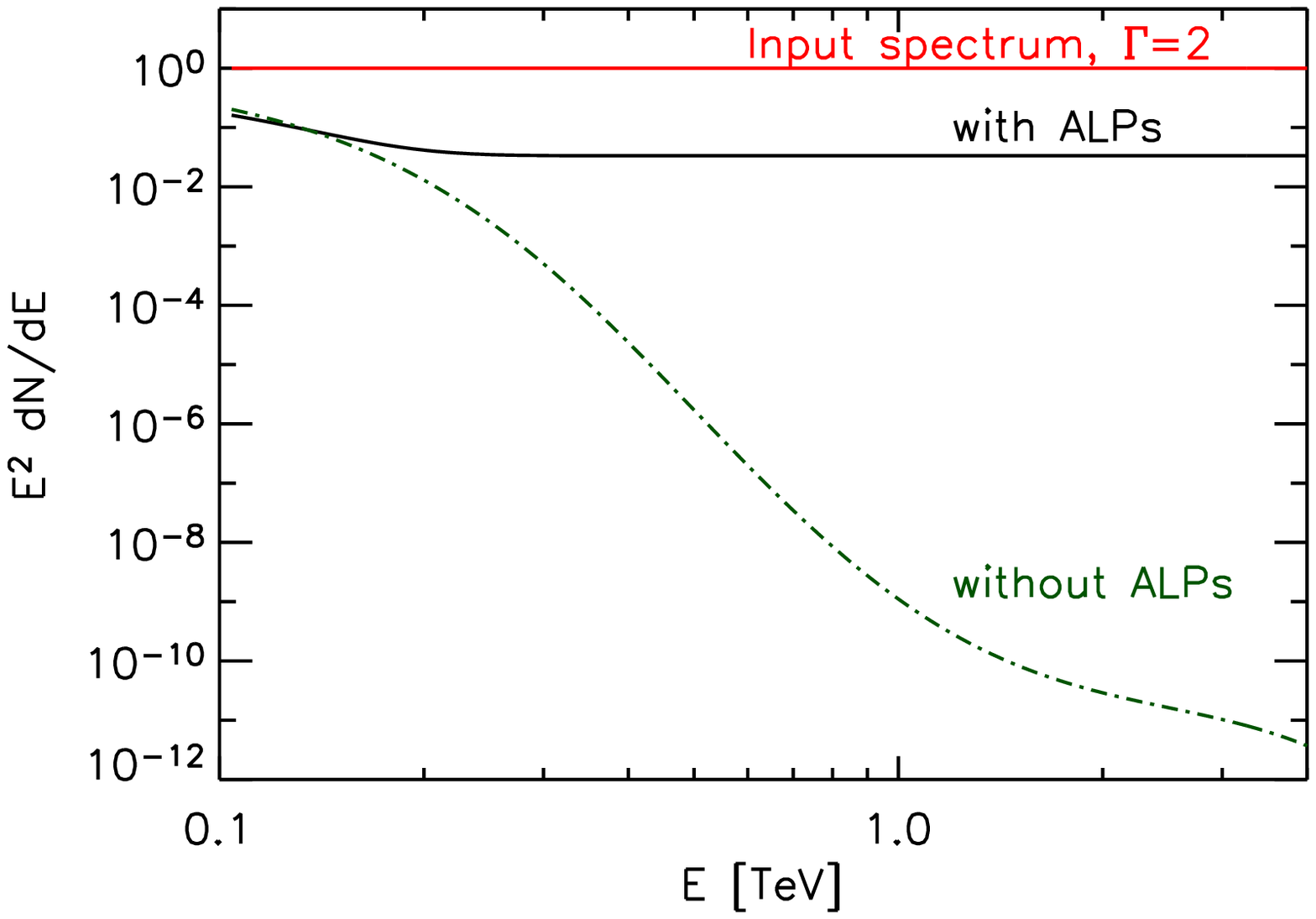,width=1.0\columnwidth, angle=0}
\end{tabular}
\caption{The gamma-ray spectrum from a source with an injected spectrum of $\d N/\d E \propto E^{-2}$, after propagation over a distance of $z=0.2$ (top) and $z=0.5$ (bottom). Results are shown with and without the effect of photon-ALP oscillations. The ALP-photon mixing mitigates the impact of absorption via pair production and leads to a plateau in the spectrum at high energies. We have calculated the effects of ALP-photon mixing assuming a source conversion probability of 0.3 and a Milky Way reconversion probability of 0.1.  The vertical axis is in arbitrary units. \label{compare}}
\end{figure}

\begin{figure}[!htb]
\begin{tabular}{c}
\epsfig{file=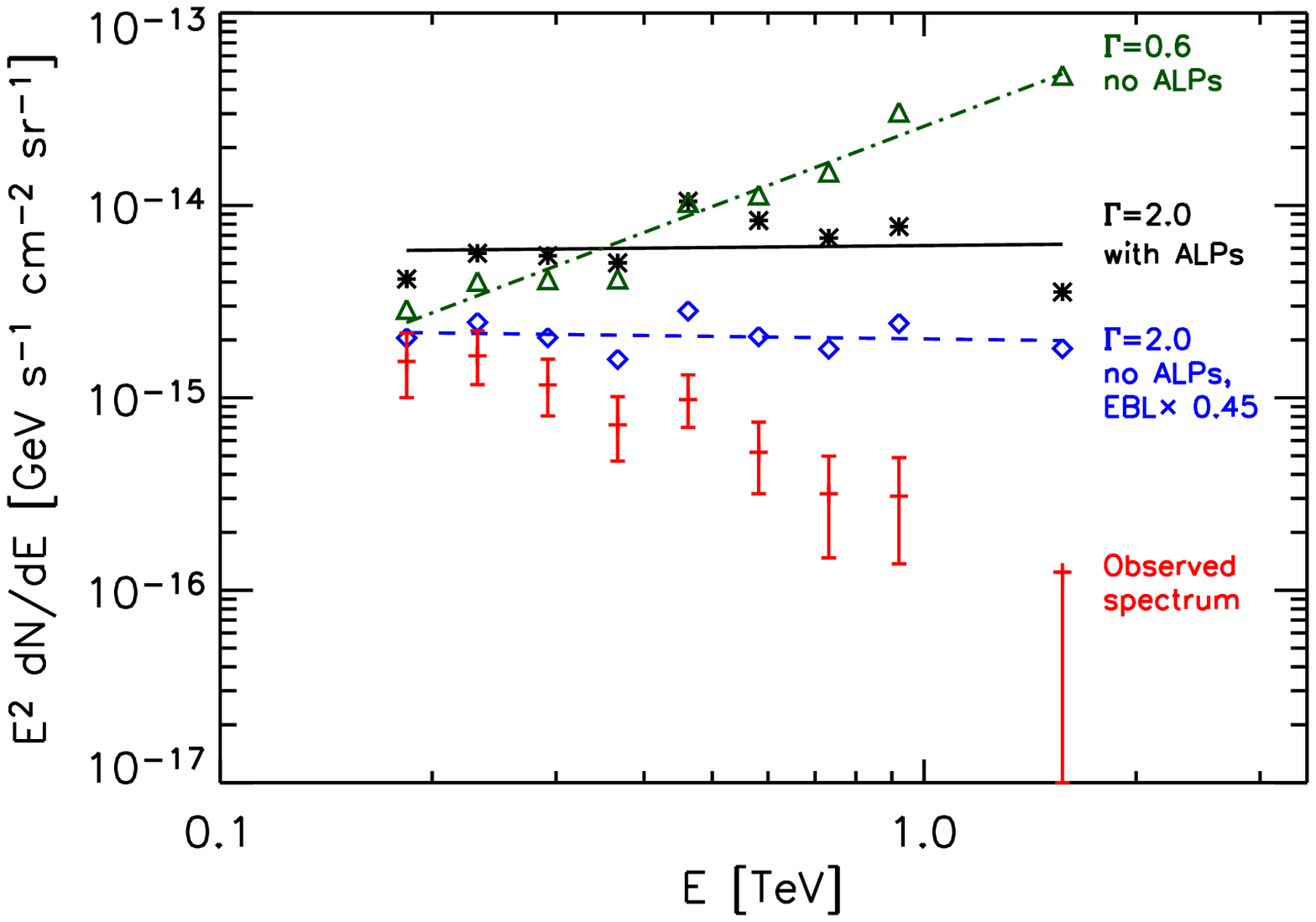,width=1.0\columnwidth, angle=0} \\
\epsfig{file=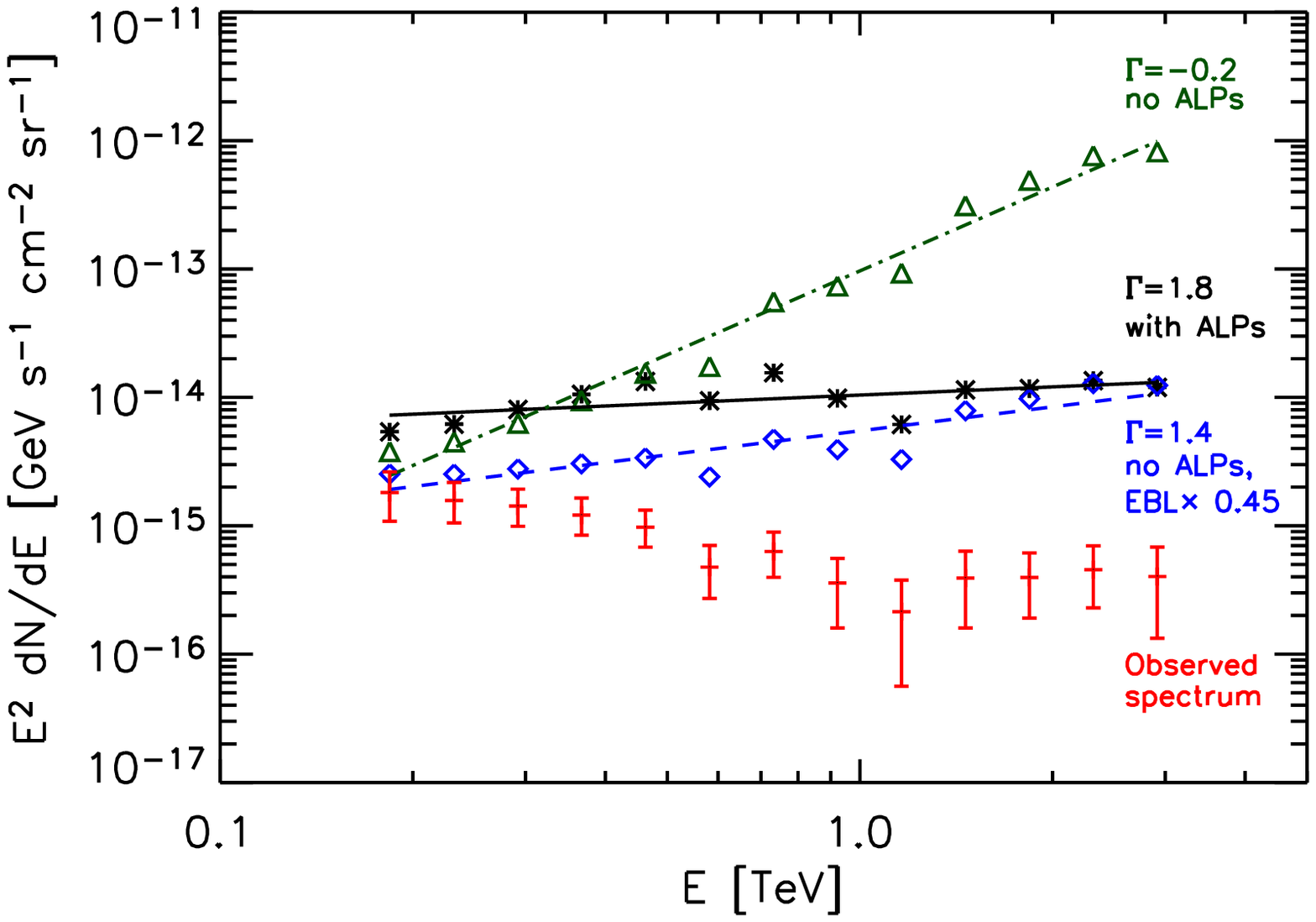,width=1.0\columnwidth, angle=0}
\end{tabular}
\caption{The points shown with error bars represent the observed gamma-ray spectrum of H 2356-309 (top) and
1ES1101-232 (bottom), from the data of Ref.~\cite{Aharonian:2005gh}. The other points shown denote the spectra required to be injected at the source in order to generate the observations after accounting for the absorption by the EBL. Using a conventional EBL spectrum and density, the reconstructed source spectra are required to be extremely hard ($\d N/ \d E\propto E^{-0.6}$ and $\d N/\d E \propto E^{0.2}$). The mechanism of photon-ALP mixing, however, can soften the required spectral slope at injection to acceptable values. See text for more details.\label{HESS}}
\end{figure}

At this point, we have all of the ingredients needed to compute the spectra of VHE gamma-ray sources, including ALP reconversion. In Fig.~\ref{compare}, we show the results for a source spectrum of $\d N/\d E \propto E^{-\Gamma}$ with $\Gamma=2$ after propagating from $z=0.2$ (top) and $z=0.5$ (bottom).  For comparison, we also include the results for a standard scenario with no ALPs. Clearly, in the presence of ALP-photon mixing, the
quasi-exponential cutoff is interrupted, and a plateau in the spectrum appears at high energies.

We next use this model to compute the source spectra of two HESS sources,
1ES1101-232 at $z=0.186$ and H 2356-309 at $z=0.165$, using the data
given in Ref.~\cite{Aharonian:2005gh}. Our results are shown as
Fig.~\ref{HESS}. The points shown with error bars represent the
spectrum as measured by HESS, whereas the other points denote the
source spectrum (before propagation) required in order to obtain the
observed spectrum. From these figures, it is clear that extremely
hard injection spectra are required ($\Gamma \approx 0.6$ and -0.2 for the two
objects, respectively) to match the observed spectrum after
propagation if one restricts oneself to the fiducial astrophysical
models for the EBL. If the effects of ALP-photon oscillations are
included, however, the required slope of the source spectrum is
softened to a reasonable value for each source ($\Gamma \approx 2.0$
and 1.8, respectively in this example, assuming $P_{a\gamma}=0.1$ in
our Galaxy). In each frame, we also show the case in which the EBL
is reduced to only 45\% of its fiducial value. This is the scenario proposed
by the HESS collaboration to explain the observed lack of
attenuation.

With this exercise, we have shown that the effects of ALP-photon mixing in known magnetic fields enables one to fit the VHE data with a reasonable spectral
index at the source ($\Gamma \approx 2$). This mechanism can accommodate the typical
predictions for the EBL without need to modify the infrared to ultraviolet emission
models from cosmic star formation, allowing spectra at the source
that are certainly much
 easier to obtain in typical acceleration models. Also, this model has several distinctive phenomenological consequences, which
we shall comment upon in the next section.

\section{Discussion and Conclusion}\label{conclusions}

In recent years, imaging atmospheric Cherenkov telescopes have
discovered many new TeV gamma-ray sources, some of which are
cosmologically distant (see Table I). These sources provide us with
a useful probe of the extragalactic background light in the infrared
to ultraviolet range. In particular, the spectra of very high energy gamma-ray
sources at cosmological distances are expected to be attenuated by
this background through the process of electron-positron pair
production. Recent observations
 seem to indicate a far greater degree of transparency of the universe to
very high energy gamma-rays than previously estimated~\cite{Aharonian:2005gh,Mazin:2007pn}. While astrophysical explanations
of these observations have been discussed~\cite{Stecker:2007jq,Stecker:2007zj}, it has also been proposed that this lack of suppression could
be the result of photon mixing with axion-like particles (ALPs), with oscillations occurring in the presence of the magnetized intergalactic medium~\cite{De Angelis:2007dy}.

In this article, we have discussed an alternative mechanism
involving photon-ALP mixing, but requiring only known magnetic
fields: namely the fields in or near the gamma-ray sources which are
needed to confine and accelerate particles (which in turn are necessary to produce gamma-rays), and the Galactic Magnetic Field of the Milky Way.  We have shown that this mechanism can be efficient for an ALP mass in the range $10^{-10}\, {\rm eV} \alt m_a\alt 10^{-8} \, {\rm eV}$ and
couplings of $g_{11}\sim 4$.  A precise
prediction of the modification expected for a given source is
precluded by our current ignorance regarding the detailed structure
of the large scale magnetic field of the Milky Way. A very
robust prediction of this model, however, (which is very different from both
the standard expectation and the model proposed in Ref.~\cite{De
Angelis:2007dy}) is that the degree of dimming observed is expected to be dependent on the galactic
coordinates $\{l,b\}$. Thus, the most striking signature would be a
reconstructed EBL density which appears to vary over different directions of the sky: consistent with standard expectations in some regions, but inconsistent in others.

The absorption of gamma-rays via pair production does not affect only point-like
source emission, but also the diffuse (or unresolved) spectrum (see
e.g. Ref.~\cite{Cuoco:2006tr}). Thus another signature of this mechanism might be an anomalous and direction-dependent spectral
behavior of diffuse radiation. This will be challenging to detect with present atmospheric Cherenkov telescopes, however, as they are not well suited for the study of diffuse radiation over large fields-of-view.  In contrast, the satellite-based experiment GLAST will observe the entire sky, although with much smaller exposures than are possible with ground based gamma-ray telescopes. Provided that the characteristic energy for the onset of
the ALP-photon conversion mechanism naturally falls in the energy range to be explored by GLAST, the
study of the diffuse radiation offers an independent test of the
crucial ingredient of this model, namely the role of the Galactic Magnetic Field. 

Through the
conversion of photons into ALPs in the Galactic Magnetic Field, a peculiar,
direction-dependent depletion of photons should arise in the
Galaxy (at up to the 30\% level). This idea was considered previously for a
different range of the ALP parameter space, impacting the diffuse X-ray
background~\cite{Krasnikov:1996bm}. The anisotropies in the diffuse
emission should anti-correlate with the regions of stronger fields as detected, for example, in Faraday rotation maps. Put another way, a peculiar
dimming of the diffuse radiation in the 1-100 GeV range
may be detected from the same regions where stronger than
expected TeV sources are present.

A detailed test of such a scenario will become increasingly possible as a
more complete picture of the high energy gamma-ray sky is developed. Since existing atmospheric Cherenkov telescopes have a
very small field of view, surveys of large fractions of the sky are
unfeasible. A targeted survey may be possible, however, as new sources in the GeV range are discovered by GLAST.
Ultimately, next generation gamma-ray observatories, such as CTA~\cite{CTA} or AGIS~\cite{AGIS}, will enable considerably more detailed studies.

The mechanism described here provides yet another example of the new opportunities for discovery made possible as a result of the recent progress in the field of high energy gamma-ray astrophysics.

\medskip
\section*{Acknowledgments}
We thank T. Weiler and D. Paneque for useful conversations. This work was supported in part by the DOE and NASA grant NAG5-10842.
Fermilab is operated by Fermi Research Alliance, LLC under Contract No.
DE-AC02-07CH11359 with the United States Department of Energy.

\end{document}